# NOVEL AND ADVANCED ACCELERATION TECHNIQUES

## CONTROL OF CHARACTERISTICS OF SELF-INJECTED AND ACCELERATED ELECTRON BUNCH IN PLASMA BY LASER PULSE SHAPING ON RADIUS, INTENSITY AND SHAPE

*V.I. Maslov[1,2], D.S. Bondar[1,2], V. Grigorencko[2], I.P. Levchuk[1], I.N. Onishchenko[1]*
[1]*National Science Center "Kharkov Institute of Physics and Technology", Kharkiv, Ukraine;*
[2]*V.N. Karazin Kharkiv National University, Kharkiv, Ukraine*

At the laser acceleration of self-injected electron bunch by plasma wakefield it is important to form bunch with small energy spread and small size. It has been shown that laser-pulse shaping on radius, intensity and shape controls characteristics of the self-injected electron bunch and provides at certain shaping small energy spread and small size of self-injected and accelerated electron bunch.
PACS: 29.17.+w; 41.75.Lx

## INTRODUCTION

The accelerating gradients in conventional linear accelerators are currently limited for technical reasons to ~100 MV/m [1], partly due to breakdown. Plasma-based accelerators have the ability to sustain accelerating gradients which are several orders of magnitude greater than that obtained in conventional accelerators [1, 2]. As plasma in experiment is inhomogeneous and nonstationary and properties of wakefield changes at increase of its amplitude it is difficult to excite wakefield resonantly by a long sequence of electron bunches (see [3, 4]), to focus sequence (see [5 - 10]), to prepare sequence from long beam (see [11 - 13]) and to provide large transformer ratio (see [14 - 20]). Providing a large transformer ratio is also being studied in dielectric accelerators (see [21 - 26]). In [4] the mechanism has been found and in [27 - 31] investigated of resonant plasma wakefield excitation by a non-resonant sequence of short electron bunches. Due to the rapid development of laser technology and physics [1, 2, 32 - 39] laser-plasma-based accelerators are of great interest now. Over the past decade, successful experiments on laser wakefield acceleration of charged particles in the plasma have confirmed the relevance of this acceleration [30 - 33, 40]. Evidently, the large accelerating gradients in the laser plasma accelerators allow to reduce the size and to cut the cost of accelerators. Another important advantage of the laser-plasma accelerators is that they can produce short electron bunches with high energy [32]. The formation of electron bunches with small energy spread was demonstrated at intense laser-plasma interactions [41]. Electron self-injection in the wake-bubble, generated by an intense laser pulse in underdense plasma, has been studied by numerical simulations [37]. Processes of a self-injection of electrons and their acceleration have been experimentally studied in a laser-plasma accelerator [42].

The problem at laser wakefield acceleration is that laser pulse quickly destroyed because of its expansion. One way to solve this problem is the use of a capillary as a waveguide for laser pulse. The second way to solve this problem is to transfer its energy to the electron bunches which as drivers accelerate witness. A transition from a laser wakefield accelerator to plasma wakefield accelerator can occur in some cases at laser-plasma interaction [43].

With newly available compact laser technology [44] one can produce 100 PW-class laser pulses with a single-cycle duration on the femtosecond timescale. With a fs intense laser one can produce a coherent X-ray pulse. Prof. T. Tajima suggested [45] utilizing these coherent X-rays to drive the acceleration of particles. Such X-rays are focusable far beyond the diffraction limit of the original laser wavelength and when injected into a crystal it interacts with a metallic-density electron plasma ideally suited for laser wakefield acceleration [45].

In [46 - 50] it has shown that at certain conditions the laser wakefield acceleration is added by a beam-plasma wakefield acceleration.

In [51] point self-injected and accelerated electron bunch was observed.

At the laser acceleration of self-injected electron bunch by plasma wakefield the accelerating gradient about 50 GV/m has been obtained in experiments [32]. In numerical simulation [50], performed according to idea of Prof. T. Tajima [45], on wakefield excitation by a X-ray laser pulse in a metallic-density electron plasma the accelerating gradient of several TV/m has been obtained. To solve the problem of laser pulse expansion, one can use a capillary discharge [52]. The second method [32, 46 - 49] of solving this problem is fast energy transfer of driver laser pulse to self-injected electron bunch. This bunch becomes driver-bunch and accelerates next self-injected electron bunch up to larger energy.

It is important to form bunch with small energy spread and small size. The purpose of this paper is to show by the numerical simulation that some laser-pulse shaping on radius, intensity and shape controls properties of the self-injected electron bunch and provides at certain conditions small energy spread and small size of self-injected and accelerated electron bunch.

## 1. PARAMETERS OF SIMULATION

Fully relativistic particle-in-cell simulation was carried out by UMKA 2D3V code [53]. A computational domain (x, y) has a rectangular shape. $\lambda$ is the laser pulse wavelength. The computational time interval is $\tau = 0.05$, the number of particles per cell is 8 and the total number of particles is $15.96 \times 10^6$. The simulation of considered case was carried out up to 800 laser periods.

The period of the laser pulse is $t_0 = 2\pi/\omega_0$. The s-polarized laser pulse enters into uniform plasma. In $y$ direction, the boundary conditions for particles, electric and magnetic fields are periodic. The critical plasma density $n_c = m_e\omega_0^2/(4\pi e^2)$. The pulse has a Gaussian profile in the transverse direction. The longitudinal and transverse dimensions of the laser pulse are selected to be shorter than the plasma wavelength. The laser pulse consists of three Gaussian micro-pulses with different properties. Each laser micro-pulse is defined with a "cos$^2$" distribution in longitudinal direction. First micro-pulse has a lower intensity and smaller radius compared to the intensity and radius of the next pulses (Figs. 1, 2). Three Gaussian micro-pulses are of different intensities and radii, so that the intensity and radius of the whole pulse grow along the pulse from first its front. Full length at half maximum of third laser pulse equals $\lambda$ and full width at half maximum equals $3\lambda$; $a_3=eE_{z0}/(m_e c\omega_0)= = 2.236$; $E_{z0}$ is the electric field amplitude. Full length at half maximum of the first pulse equals $4\lambda$ and full width at half maximum equals $\lambda$, $a_{01}=eE_{z01}/(m_e c\omega_0) = 1$. Full length at half maximum of the second pulse equals $2\lambda$ and full width at half maximum equals $2\lambda$, $a_{02}=eE_{z02}/(m_e c\omega_0) = 1.732$. Coordinates $x$ and $y$, time $t$, electric field amplitude $E_z$ and electron plasma density $n_0$ are given in dimensionless form in units of $\lambda$, $2\pi/\omega_0$, $m_e c\omega_0/(2\pi e)$, $m_e\omega_0^2/(16\pi^3 e^2)$.

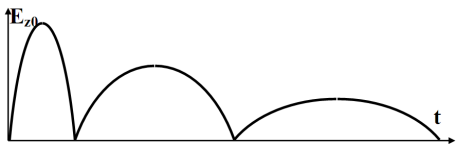

*Fig. 1. Laser pulse, shaped on intensity*

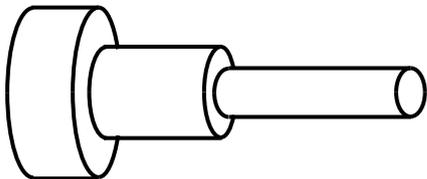

*Fig. 2. Laser pulse, shaped on radius and shape*

## 2. RESULTS OF SIMULATION

We consider the wakefield excitation by laser pulse, shaped on intensity and radius. The intensity and radius grow from the first front of the pulse and then abruptly break off. Let us show that one can control the wakefield using laser pulse shaping on radius, intensity and shape.

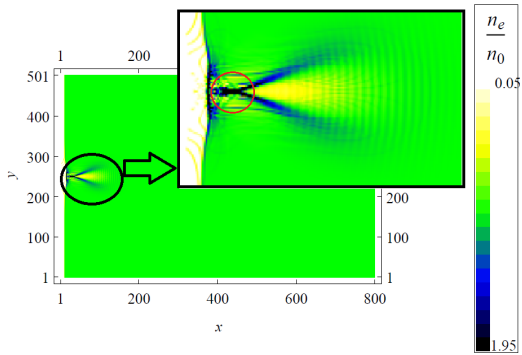

*Fig. 3. Wake perturbation of plasma electron density $n_e$ excited by laser pulse near the boundary of injection*

Namely, stochastization of the wakefield is suppressed due to the adiabatic growth of intensity and radius of the pulse. Also one can control quality of the bunch. Namely, a point-kind bunch is formed.

From Figs. 3, 4 one can see that due to selected spatial distribution of intensity and radius of laser pulse near the boundary of injection the adiabatic structure, suitable for good self-injection, is formed.

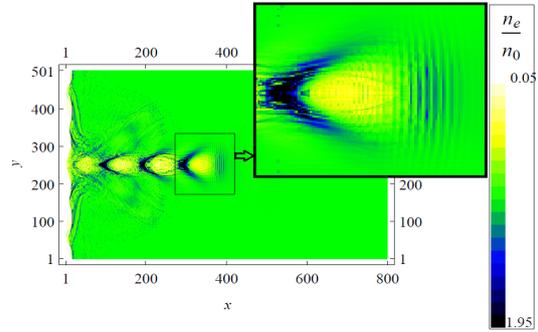

*Fig. 4. Electron bunch self-injection at plasma wakefield excitation by laser pulse, shaped on radius and intensity*

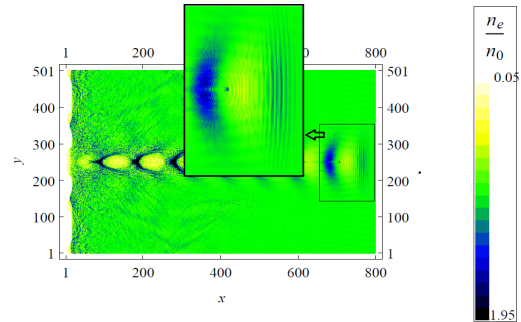

*Fig. 5. Self-injected electron bunch acceleration by wakefield bubble, excited in plasma by laser pulse, shaped on radius, intensity and shape*

It is seen from Fig. 4 that short electron bunch is self-injected. Fig. 5 demonstrates that point-kind electron bunch is accelerated.

## CONCLUSIONS

Thus, the authors for the first time used laser pulse shaping on radius, intensity and shape to control the parameters of a self-injected and accelerated electron bunch. Previously, no one used laser pulse shaping on radius, intensity and shape to control the parameters of a self-injected and accelerated electron bunch. It has been shown by the numerical simulation that laser pulse shaping on radius, intensity and shape controls properties of the self-injected electron bunch and provides at certain conditions small energy spread and small size of self-injected and accelerated electron bunch.

## УПРАВЛЕНИЕ ХАРАКТЕРИСТИКАМИ САМОИНЖЕКТИРОВАННОГО И УСКОРЯЕМОГО ЭЛЕКТРОННОГО СГУСТКА В ПЛАЗМЕ ПРОФИЛИРОВАНИЕМ ЛАЗЕРНОГО ИМПУЛЬСА ПО РАДИУСУ, ИНТЕНСИВНОСТИ И ФОРМЕ

*В.И. Маслов, Д.С. Бондарь, В. Григоренко, И.П. Левчук, И.Н. Онищенко*

При лазерном ускорении самоинжектированного сгустка электронов плазменным кильватерным полем важно сформировать сгусток с небольшим разбросом по энергии и небольшим размером. Показано, что профилирование лазерного импульса по радиусу, интенсивности и форме контролирует характеристики самоинжектированного электронного сгустка и обеспечивает при определенном профилировании небольшой энергетический разброс и малый размер сгустка электронов.

## УПРАВЛІННЯ ХАРАКТЕРИСТИКАМИ САМОІНЖЕКТОВАНОГО І ПРИСКОРЕНОГО ЕЛЕКТРОННОГО ЗГУСТКА В ПЛАЗМІ ПРОФІЛЮВАННЯМ ЛАЗЕРНОГО ІМПУЛЬСУ ПО РАДІУСУ, ІНТЕНСИВНОСТІ І ФОРМІ

*В.І. Маслов, Д.С. Бондарь, В. Грігоренко, І.П. Левчук, І.М. Оніщенко*

При лазерному прискоренні самоінжектованого згустку електронів плазмовим кільватерним полем важливо сформувати згусток з невеликим розкидом енергії і невеликим розміром. Показано, що профілювання лазерного імпульсу по радіусу, інтенсивності та формі контролює характеристики самоінжектованого електронного згустку і забезпечує при певному профілюванні невеликий енергетичний розкид і малий розмір самоінжектованого і прискореного згустку електронів.